\documentclass[aps,pra,twocolumn,showpacs,floatfix]{revtex4}
\usepackage{epsfig}
\usepackage{graphicx}
\usepackage{dcolumn}
\usepackage{amsthm,amsmath}

\begin{document}
\title{Van der Waals Coefficients for the Alkali-metal Atoms in the Material Mediums}
\author{$^a$Bindiya Arora \footnote{Email: arorabindiya@gmail.com} and $^b$B. K. Sahoo  \footnote{Email: bijaya@prl.res.in}}
\affiliation{$^a$Department of Physics, Guru Nanak Dev University, Amritsar, Punjab-143005, India,\\
$^b$Theoretical Physics Division, Physical Research Laboratory, Navrangpura, Ahmedabad-380009, India}
\date{Received date; Accepted date}
 
\begin{abstract}
The damping coefficients for the alkali atoms are determined very accurately by
taking into account the optical properties of the atoms and three distinct types
of trapping materials such as Au (metal), Si (semi-conductor) 
and vitreous SiO$_2$ (dielectric). Dynamic dipole polarizabilities are calculated
precisely for the alkali atoms that reproduce the damping coefficients 
in the perfect conducting medium within 0.2\% accuracy. Upon the consideration of
the available optical data of the above wall materials, the damping
coefficients are found to be substantially different than those of the ideal conductor.
We also evaluated dispersion coefficients for the alkali dimers and compared them 
with the previously reported values. These coefficients are fitted into a ready-to-use 
functional form to aid the experimentalists the interaction potentials only with the 
knowledge of distances.
\end{abstract}

\maketitle

Accurate information on the long-range interactions 
such as dispersion (van der Waals) and retarded (Casimir-Polder) potentials
between two atoms and between an atom and surface of the trapping material are necessary for the 
investigation of the underlying physics of atomic collisions especially in the ultracold atomic 
experiments \cite{lifshitzbook,casimir1,london,israel}. Presence of atom-surface interactions lead to 
a shift in the oscillation frequency of the trap which alters the trapping frequency as well 
as magic wavelengths for state-insensitive trapping of the trapped condensate. 
Moreover, this effect has also gained interest in generating novel atom optical devices known as the ``atom chips".
In addition,
the knowledge of dispersion coefficients is required in experiments of photo-association, 
fluorescence spectroscopy, determination of scattering lengths, analysis  of feshbach resonances, 
determination of stability of Bose-Einstein condensates (BECs), probing extra dimensions to accommodate
Newtonian gravity in quantum mechanics etc. \cite{roberts,amiot,leo,harber,leanhardt,lin}.   

There have been many experimental evidences of an attractive force between neutral atoms and between 
neutral atoms with trapping surfaces but their precise determinations are relatively difficult. 
In the past two decades, several groups have evaluated dispersion coefficients $C_3$ defining interaction 
between an atom and a wall using various approaches \cite{derevianko1,derevianko4,jiang}
without rigorous estimate of uncertainties. More importantly, they are evaluated for a perfect conducting wall
which are quite different from an actual trapping wall. Since these coefficients depend on the dielectric constants 
of the materials of the wall, therefore it is worth determining them precisely for trapping materials with varying 
dielectric constants (for good conducting, semi conducting, and dielectric mediums) as has been attempted in
\cite{lach1,caride}. Casimir and Polder 
\cite{casimir1} had estimated that at intermediately large separations the retardation effects 
of the virtual photons passing between the atom and its image weakens the attractive atom-wall force and 
the force scales with a different power law (given in details below). In this paper, we carefully examine these retardation or
damping effects which have not been extensively studied earlier. We also parameterized our damping coefficients into a readily usable form to be used in experiments.

The atom-surface interaction potential resulting from the fluctuating dipole moment of an atom 
interacting with its image in the surface is formulated by  \cite{lifshitzbook,lach1}
{\small \begin{equation}
U^a(R)= -\frac{\alpha_{fs}^3}{2\pi}\int_0^{\infty}d\omega \omega^3\alpha(\iota\omega)\int_1^{\infty}d\xi e^{-2\alpha_{fs}\xi\omega R} H(\xi,\epsilon(\iota\omega)),
\label{atwp}
\end{equation}}
where $\alpha_{fs}$ is the fine structure constant, $\epsilon(\omega)$ is the frequency dependent dielectric constant of 
the solid, $R$ is the distance between the atom and the surface and $\alpha(\iota\omega)$ is the ground state 
dynamic polarizability with imaginary argument. The function $H(\xi,\epsilon(\iota\omega))$ 
is given by
\begin{equation}
H(\xi,\epsilon)=(1-2\xi^2)\frac{\sqrt{\xi^2+\epsilon-1}-\epsilon\xi}{\sqrt{\xi^2+\epsilon-1}+\epsilon\xi} + \frac{\sqrt{\xi^2+\epsilon-1}-\xi}{\sqrt{\xi^2+\epsilon-1}+\xi}\nonumber
\end{equation}
with the Matsubara frequencies denoted by $\xi$.

\begin{figure}[t]
 \includegraphics[width=8.5cm,height=6.5cm]{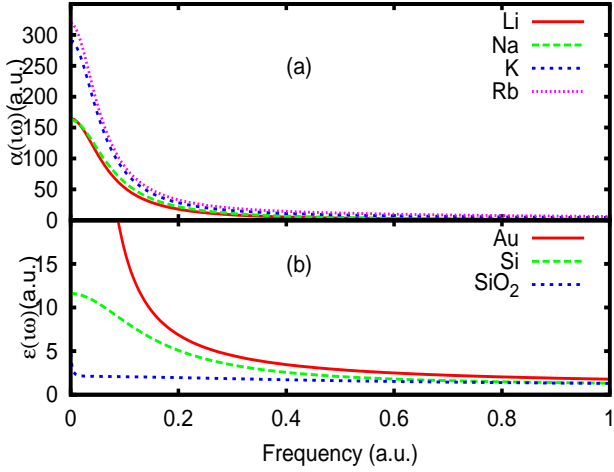}
  \caption{Dynamic polarizabilities of the Li, Na, K and Rb atoms and dielectric permittivity of the Au, Si and 
  SiO$_2$ surfaces along the imaginary axis as functions of frequencies.}
  \label{pol}     
\end{figure}

In asymptotic regimes, the Matsubara integration is dominated by its first term and the 
potential can be approximated to $U^a(R)=-\frac{C_{3T}}{R^3}$ with $C_{3T}=\frac{\alpha(0)}{4} 
(\epsilon(0)-1)/(\epsilon(0)+1)$. The potential form can be described more accurately at the retardation distances as
$U^a(R)=-\frac{C_4}{R^4}$ and at the non-retarded region as $U^a(R)=-\frac{C_3}{R^3}$ \cite{casimir1}. To express the 
potential in the intermediate region, these approximations are usually modified either to 
$U^a(R)=-\frac{C_4}{(R+\lambdabar)R^3}$ or to $U^a(R)=-\frac{C_3}{R^3} f_3(R)$ where $\lambdabar$ and $f_3(R)$ are respectively
known as the reduced wavelength and damping function. It would be interesting to testify the validity 
of both the approximations by evaluating $C_3$, $C_4$ and  $f_3(R)$ coefficients together for different atoms in 
conducting, semi-conducting and dielectric materials. Since the knowledge of magnetic permeability of the 
material is required to evaluate $C_4$ coefficients, hence we determine only the $C_3$ and $f_3(R)$ coefficients. 
With the knowledge of $C_3$ and $f_3$ values, the atom-surface interaction potentials can be easily reproduced
and they can be generalized to other surfaces.
\begin{table}[t]
\caption{\label{c3} Calculated $C_3$ coefficients along with their uncertainties for the alkali-metal atoms and their comparison with other reported values. 
Classification of various contributions are in accordance with  \cite{derevianko1}$^a$,  \cite{derevianko4}$^b$ and  \cite{kharchenko}$^c$.}
\begin{ruledtabular}
\begin{tabular}{lcccc}
     		&  Li       &  Na          & K         & Rb     \\
\hline
\multicolumn{4}{c}{Perfect Conductor}  \\	
Core            & 0.074     &    0.332     	& 0.989     & 1.513   \\
Valence 		& 1.387     &    1.566     	& 2.115     & 2.254    \\
Core-Valence 	& $\sim 0$  & $\sim 0$ 		&$-0.016$   &$-0.028$  \\
Tail           	& 0.055     & 0.005			& 0.003	    &0.003      \\
&  &  \\
Total	  		& 1.516(2)	    		& 1.904(2)			&3.090(4)	    		&3.742(5)\\
Others			& 1.5178 $^{\rm a}$ & 1.8858$^{\rm b}$ 	& 2.860$^{\rm b}$ 	&3.362$^{\rm b}$ \\
				& & 1.889$^{\rm c}$	& \\
\multicolumn{4}{c}{Metal: Au}                     \\	
Core            & 0.010     & 0.051     & 0.263     &  0.419  \\
Valence 		& 1.160     & 1.285     & 1.804     &  1.927   \\
Core-Valence 	& $\sim 0$  & $\sim 0$ 	& -0.005    &  -0.010\\
Tail           	&  0.029    & 0.002		& 0.001	    &  0.002    \\
&  &  \\
Total  			& 1.199(2)	    & 1.338(1)	    & 2.062(4)		& 2.338(4)   \\
Others~\cite{lach1}& 1.210 	& 1.356  & 2.058 & 2.79 \\
\multicolumn{4}{c}{Semi-conductor: Si}                      \\	
Core            &  0.006    &  0.033    & 0.184     & 0.299   \\
Valence 		&  0.993    &  1.099    & 1.543     & 1.649    \\
Core-Valence 	& $\sim 0$  & $\sim 0$ 	& -0.004 	&-0.008 \\
Tail           	&  0.023    & 0.002		& 0.001	    & 0.001     \\
&  &  \\
Total 			& 1.022(2)	   & 1.134(1) 		& 1.724(3) & 1.942(4)   \\
\multicolumn{4}{c}{Dielectric: SiO$_2$}                      \\	
Core            &  0.004    & 0.022        	& 0.116     & 0.184   \\
Valence 		&  0.468    &  0.519       	& 0.726     & 0.775    \\
Core-Valence 	& $\sim 0$  & $\sim 0$ 		& -0.002 	& -0.004 \\
Tail           	&  0.012    & 0.001			& 	0.001   & 0.001     \\
&  &  \\
Total          	& 0.4844(8)  	& 0.5424(5)			& 0.839(1)		&  0.956(2) \\
\end{tabular} 
\end{ruledtabular}
\end{table}
In general, the $C_3$ coefficient is given by
\begin{equation}
C_3 \approx \frac{1}{4 \pi} \int_0^{\infty} d \omega \alpha(\iota\omega) \frac{\epsilon(\iota\omega)-1}{\epsilon(\iota\omega)+1}.
\label{c3eq}
\end{equation}
For a perfect conductor $\epsilon \rightarrow \infty$, $\frac{\epsilon(\iota\omega)-1}{\epsilon(\iota\omega)+1} \rightarrow 1$ and 
for other materials with their refractive indices $n = \sqrt{\epsilon}$ varying between 1 and 2, 
$\frac{\epsilon(\iota\omega)-1}{\epsilon(\iota\omega)+1} \approx \frac{\epsilon(0)-1}{\epsilon(0)+1}$ is nearly a constant and
can be approximated to 0.77. For more preciseness, it is necessary to consider the actual frequency dependencies
of $\epsilon$s in the materials. In the present work, three distinct materials such as Au, Si and SiO$_2$ belonging to 
conducting, semi-conducting and dielectric objects respectively, are taken into account to find out $f_3(R)$ functions 
and compared against a perfect conducting wall for which case we express \cite{kharchenko}
\begin{eqnarray}
f_3(R)&=&\frac{1}{4\pi C_3}\int_0^{\infty}d\omega \alpha(\iota\omega) e^{-2 \alpha_{fs} \omega R} Q(\alpha_{fs}\omega R)
\label{f3eqp},
\end{eqnarray}
with $Q(x)=2x^2+2x+1$. To find out $f_3(R)$ for the other surfaces, we evaluate $U^a(R)$ by substituting 
their $\epsilon(\iota \omega)$ values in Eq. (\ref{atwp}). 
  
\begin{figure}[t]
 \includegraphics[width=8.7cm,height=7.0cm]{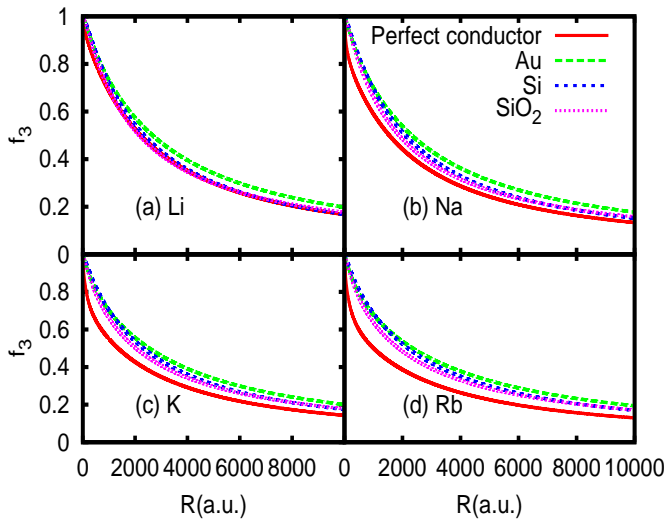}
  \caption{The retardation coefficient $f_3$(R) (dimensionless) for Li, Na, K and Rb as a function of atom-wall distance R. }
  \label{f3fg}     
\end{figure}
Similarly, the leading term in the long-range interaction between two atoms denoted by $a$ and $b$ is
approximated by $U^{ab}(R) = -\frac{C_6^{ab}}{R^6}$, where the $C_6^{ab}$ is known as the van der Waals coefficient and 
$R$ is the  distance between two atoms.  If retardation effects are included then it is modified to
$U^{ab}(R)=-\frac{C_6^{ab}}{R^6}f_6^{ab}(R)$.
The dispersion coefficient $C_6^{ab}$ and the damping coefficient $f_6^{ab}(R)$ between the atoms can be 
estimated using the expressions \cite{kharchenko}
\begin{eqnarray}
C_6^{ab} &=&\frac{3}{\pi}\int_0^{\infty}d\omega \alpha^a(\iota\omega)\alpha^b(\iota\omega), \ \ \ \ \text{and} \nonumber\\
f_6^{ab} &=&\frac{1}{\pi C_6^{ab}}\int_0^{\infty}  d\omega  \alpha^a(\iota\omega)\alpha^b(\iota\omega) e^{-2\alpha_{fs}\omega R}P(\alpha_{fs}\omega R),\nonumber
\end{eqnarray}
where $P(x)=x^4+2x^3+5x^2+6x+3$. 

\begin{table}[t]
\caption{\label{f3} Fitting parameters $a$ and $b$ for $f_3$ coefficients with a perfectly conducting wall, Au, Si, and SiO$_2$ surfaces.}
\begin{ruledtabular}
\begin{tabular}{lcccc}
  & Li & Na & K & Rb\\
\hline
{Perfect Conductor}  \\	
a 		& 0.9843 & 1.0802 & 1.1845 & 1.2598 \\
b 		& 0.0676 & 0.0866 & 0.0808 & 0.0907 \\
{Metal: Au}                     \\
a 		& 0.9775 & 0.9846 & 1.0248 & 1.0437 \\
b 		& 0.0675 & 0.0614 & 0.0532 & 0.0558 \\
{Semi-conductor: Si}                      \\	
a 		& 0.9436 & 0.9436 & 0.9749 & 0.9869 \\
b 		& 0.0638 & 0.0718 & 0.0622 & 0.0647 \\
{Dielectric: SiO$_2$}                      \\	
a 		& 0.9754 & 0.9789 & 1.0238 & 1.0423 \\
b 		& 0.0650 & 0.0746 & 0.0649 & 0.0685 \\
\end{tabular} 
\end{ruledtabular}
\end{table}
Using our previously reported E1 matrix elements \cite{arora-sahoo1,arora-sahoo2} and experimental energies, we plot the 
dynamic polarizabilities of the ground states in Fig. \ref{pol} of the considered alkali atoms. The static polarizabilities
corresponding to $\omega=0$ come out to be 164.1(7), 162.3(2), 289.7(6) and 318.5(8), as given in \cite{arora-sahoo1,
arora-sahoo2}, against the experimental values 
164.2(11) \cite{li-exp}, 162.4(2) \cite{na-exp}, 290.58(1.42) \cite{k-exp} and 318.79(1.42) \cite{k-exp} 
in atomic unit (a.u.) for Li, Na, K and Rb atoms respectively. 
It clearly indicates the preciseness of our estimated results. The main reason for achieving such high accuracies
in the estimated static polarizabilities is due to the use of E1 matrix elements extracted from the
precise  lifetime measurements of few excited states and by fitting our E1 results obtained from the relativistic coupled-cluster calculation
at the singles, doubles and partial triples excitation level (CCSD(T) method) to the measurements of the static 
polarizabilities of the excited states. 

\begin{table*}[t]
\caption{\label{c6} $C_6$ coefficients with fitting parameters for the alkali dimers. Contributions from the valence,
core and valence-core polarizabilities alone are labeled as $C_6^v$, $C_6^c$ and $C_6^{vc}$, respectively 
and $C_6^{ct}$ corresponds to contributions from the remaining cross terms.  References: $^a$\cite{derevianko5}, $^b$\cite{marinescu}, $^c$\cite{mitroy1}, $^d$\cite{sahooli}, $^e$\cite{russier}, $^f$\cite{pashov}, $^g$\cite{chin}.}
\begin{ruledtabular}
\begin{tabular}{lccccccccc}
\multicolumn{1}{l}{Dimer} &
\multicolumn{1}{c}{$C_6^v$} &
\multicolumn{1}{c}{$C_6^c$} &
\multicolumn{1}{c}{$C_6^{vc}$} &
\multicolumn{1}{c}{$C_6^{ct}$} &
\multicolumn{1}{c}{$C_6$(Total})&
\multicolumn{1}{c}{Others}&
\multicolumn{1}{c}{Exp}&
\multicolumn{1}{c}{a}&
\multicolumn{1}{c}{b}\\
\hline
Li-Li & 1351 & 0.07   & $\sim$ 0 & 39    & 1390(4)   & 1389(2)$^{\rm a}$,1388$^{\rm b}$,1394.6$^{\rm c}$ ,1473$^{\rm d}$ &	& 0.8592 & 0.0230\\
Li-Na & 1428 & 0.32   & $\sim$ 0 & 37    & 1465(3)   & 1467(2)$^{\rm a}$&  & 0.8592 & 0.0245 \\
Li-K  & 2201 & 1.27   & $\sim$ 0 & 119   & 2321(6)   & 2322(5)$^{\rm a}$&  & 0.8640 & 0.0217 \\
Li-Rb & 2368 &1.94    & $\sim$ 0 & 179   & 2550(6)  &2545(7)$^{\rm a}$ &  & 0.8666 & 0.0262 \\
& & & & & & &\\
Na-Na & 1515 &  1.51  & $\sim$ 0 & 33    & 1550(3)   & 1556(4)$^{\rm a}$, 1472$^{\rm b}$,1561$^{\rm c}$& & 0.8591 & 0.0262 \\
Na-K  & 2316 & 6.24   & $\sim$ 0 & 118   & 2441(5)   & 2447(6)$^{\rm a}$& 2519$^{\rm e}$ & 0.8555 & 0.0231 \\
Na-Rb & 2490 & 9.60   & $\sim$ 0 & 184   & 2684(6)  & 2683(7)$^{\rm a}$&	&	0.8686 & 0.0232 \\
& & & & & & &\\
K-K   & 3604 &  29.89 & 0.01     & 261   & 3895(15)  & 3897(15)$^{\rm a}$, 3813$^{\rm b}$,3905$^{\rm c}$ &3921$^{\rm f}$ & 0.8738 & 0.0207 \\
K-Rb  & 3880 &  46.91 & 0.02     & 465   & 4384(12)  & 4274(13)$^{\rm a}$ &	& 0.8738 & 0.0207 \\
& & & & & && \\
Rb-Rb &4178  & 73.96  & 0.4      & 465   & 4717(19)  & 4691(23)$^{\rm a}$, 4426$^{\rm b}$,4635$^{\rm c}$& 4698$^{\rm g}$ & 0.8779 & 0.0207 \\
\end{tabular} 
\end{ruledtabular}
\end{table*}
Substituting the dynamic polarizabilities in Eq. (\ref{c3eq}), we evaluate the $C_3$ coefficients for a perfect conductor 
(to compare with previous studies), for a real metal Au, for a semi conductor object Si and for a dielectric
substance of glassy structure SiO$_2$. These values are given in Table \ref{c3}
with break down from various individual contributions and estimated uncertainties are quoted in the parentheses after ignoring errors from the used experimental 
data. To achieve the claimed accuracy in our results it was necessary to use the complete tabulated 
data for the refraction indices of Au, Si, and SiO$_2$ to calculate their dielectric permittivities
at all the imaginary frequencies \cite{palik}. We evaluate the imaginary parts of the dielectric constants using the relation
$\rm{Im} \left (\epsilon (\omega) \right ) = 2n (\omega)\kappa(\omega)$,
where n and $\kappa$ are the real and imaginary parts of the refractive index of a material. The available data for 
Si and SiO$_2$ are sufficiently extended to lower frequencies. However, they are extended to the lower 
frequencies for Au with the help of the Drude dielectric function~\cite{caride}
\begin{equation}
\epsilon(\omega) = 1-\frac{\omega_p^2}{\omega(\omega+\iota\gamma)},
\end{equation}
with relaxation frequency $\gamma=0.035$ eV and plasma frequency $\omega_p = 9.02$ eV. The corresponding real values 
at imaginary frequencies are obtained by using the Kramers-Kronig formula
\begin{equation}
\rm{Re}(\epsilon(\iota\omega))=1+\frac{2}{\pi}\int_0^{\infty}d\omega' \frac{\omega' \rm{Im}(\epsilon(\omega'))}{\omega^2+{\omega'}^2}.\label{eq-kk}
\end{equation}
In bottom part of Fig.~\ref{pol}, the $\epsilon (\iota\omega)$ values as a function of imaginary frequency are 
plotted for Au, Si, and SiO$_2$. The behavior of $\epsilon (\iota\omega)$ for various materials is obtained as expected and they match well with the graphical representations given by Caride and co-workers \cite{caride}.

As shown in Table~\ref{c3}, $C_3$ coefficients increase with the increase in atomic mass. First we present our results for the $C_3$ coefficients for the interaction of these atoms with a perfectly conducting wall. The dominant contribution to the $C_3$ coefficients is from the 
valence part of the polarizability. We also observed that the core contribution to the $C_3$ 
coefficients increases with the  increasing number of electrons in the atom which is in agreement with the 
prediction made in Ref. \cite{derevianko4}. Our results are also in good agreement with the results reported by 
Kharchenko \textit{et al.} \cite{kharchenko} for Na. Therefore, our results obtained for other materials
seem to be reliable enough.  We noticed that the $C_3$ coefficients for a perfect conductor were approximately 1.5, 2, and 3.5 times larger than the $C_3$ coefficients for Au, Si, and SiO$_2$ respectively. The decrease in the coefficient values for the considered mediums  can be attributed to the fact that in case of dielectric material the theory is modified for non-unity reflection and for different origin of the transmitted waves from the surface. In addition to this, for Si and SiO$_2$ there are additional interactions due to charge dangling bonds specially at shorter separations.
The recent estimations with Au medium carried out by Lach \textit{et al.}~\cite{lach1}
are in agreement with our results since the polarizability database they have used is taken from Ref.
\cite{derevianko4}. These calculations seem to be sensitive on the choice of grids used for the numerical 
integration. An exponential grid yield the results more accurately and it is insensitive to choice of the 
size of the grid in contrast to a linear grid. In fact with the use of a linear grid having a spacing
$0.1$, we observed a 3-5\% fall in $C_3$ coefficients for the considered atoms. The reason being that
most of the contributions to the evaluation of these coefficients come from the lower frequencies 
which yield inaccuracy in the results for large grid size.

Fig.~\ref{f3fg} shows a comparison of the $f_3(\text{R})$ values obtained for Li, Na, K, and Rb atoms as a function 
of atom-wall separation distance R for the four different materials studied in this work. As seen in the figure, 
the retardation coefficients are the smallest for an ideal metal. At very short separation distances the results 
for a  perfectly conducting material differs from the results of Au, Si, and SiO$_2$ by less than 4\%. As the 
atom-surface distance increases, the deviations of $f_3$ results for various materials from the results of an
ideal metal are considerable and vary as 18\%, 15\% and 6\% for Li; 33\%, 14\% and 18\% for Na; 40\%, 13\% 
and 26\% for K; and 50\%, 13\% and 33\% for Rb in  Au, Si, and SiO$_2$ surfaces respectively. The deviation of
results between an ideal metal and other dielectric surfaces is smallest for the Li atom and increases appreciably 
for the Rb atom. 
We use the functional form to describe 
accurately the atom-wall interaction potential at the separate distance R as
\begin{eqnarray}
f_3(R) = \frac{1}{a+b(\alpha_{fs}R)} \label{fit}.
\end{eqnarray}
By extrapolating data from the above figure, we list the extracted $a$ and $b$ values for the considered atoms in
all the materials in Table \ref{f3}.

\begin{figure}[t]
 \includegraphics[width=8.7cm,height=5.0cm]{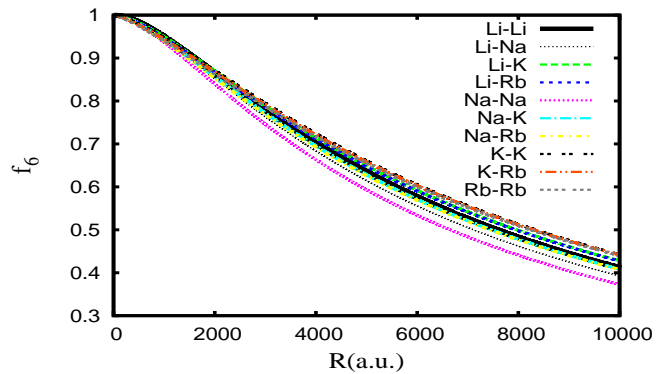}
  \caption{The retardation coefficient $f_6$(R) (dimensionless) for the alkali dimers as a function of atom-atom distance R. }
  \label{fig-f6}     
\end{figure}

In Table \ref{c6}, we present our calculated results for the $C_6$ coefficients for the alkali dimers. 
In columns II, III and IV, we give individual contributions from the valence, core and valence-core 
polarizabilities to $C_6$ evaluation and column V represents contributions from the cross terms which
are found to be crucial for obtaining accurate results. As can be seen from Table \ref{c3}, the trends 
are almost similar to $C_3$ evaluation. A comparison of our $C_6$ values  
with other recent calculations and available experimental results is also presented in the same table.  
Using the similar fitting procedure 
as for $f_3$, we obtained fitting parameters $a$ and $b$ for $f_6$ from Fig.~\ref{fig-f6} which are quoted 
in the last two columns of the above table. 

To summarize, we have investigated the dispersion and damping coefficients for the atom-wall and atom-atom 
interactions for the Li, Na, K, and Rb atoms and their dimers in this work. The interaction potentials of the alkali atoms are studied 
with Au, Si, and SiO$_2$ surfaces and found to be very different than a perfect conductor. It is also
shown that the interaction of the atoms in these surfaces is considerably distinct from each other.
A readily usable functional form of the retardation coefficients for the interaction between two alkali atoms 
and alkali atom with the above mediums is provided. Our fit explains more than 99\% of total variation in data about
average. The results are compared with the other theoretical and experimental values.

 The work of B.A. is supported by the CSIR, India (Grant no. 3649/NS-EMRII). We thank Dr. G. Klimchitskaya and Dr. G. Lach for some useful discussions. B.A. also thanks Mr. S. Sokhal for his help in some calculations. Computations were carried 
out using 3TFLOP HPC Cluster at Physical Research Laboratory, Ahmedabad.

\end{document}